\documentclass{vldb}
\usepackage{graphicx}
\usepackage{balance}
\usepackage{latexsym} 
\usepackage{listings}
\usepackage{todonotes}
\usepackage{enumitem}
\usepackage{hyperref}
\usepackage{subfig}
\usepackage{multirow}
\usepackage{algorithm}
\usepackage{amsmath}
\usepackage{colortbl}
\usepackage{algpseudocode}

\usepackage{tikz}
\usetikzlibrary{arrows}

\newtheorem{exam}{Example}

\newcommand{\mypar}[1]{\smallskip\noindent\textbf{#1}.}
\newcommand{\myparr}[1]{\noindent\textbf{#1}.}

\DeclareRobustCommand\fignumref[1]{\raisebox{.5pt}{\textcircled{\raisebox{-.75pt} {\small #1}}}}

\definecolor{black}{rgb}{0,0,0}
\definecolor{grey}{rgb}{0.8,0.8,0.8}
\definecolor{red}{rgb}{1,0,0}
\definecolor{green}{rgb}{0,1,0}
\definecolor{darkgreen}{rgb}{0,0.5,0}
\definecolor{darkpurple}{rgb}{0.5,0,0.5}
\definecolor{darkdarkpurple}{rgb}{0.3,0,0.3}
\definecolor{blue}{rgb}{0,0,1}
\definecolor{shadegreen}{rgb}{0.95,1,0.95}
\definecolor{shadeblue}{rgb}{0.95,0.95,1}
\definecolor{shadered}{rgb}{1,0.85,0.85}
\definecolor{shadegrey}{rgb}{0.85,0.85,0.85}
\definecolor{oddRowGrey}{rgb}{0.80,0.80,0.80}
\definecolor{evenRowGrey}{rgb}{0.85,0.85,0.85}

\newcommand{\thead}[1]{{\cellcolor{black}{\textcolor{white}{\textbf{#1}}}}}

\begin{document}
\definecolor{Comments}{rgb}{0.00,0.50,0.00}
\definecolor{KeyWords}{rgb}{0.00,0.00,0.13}
\definecolor{Strings}{rgb}{0.60,0.00,0.00}

\lstdefinestyle{psql}
{
  tabsize=3,
  basicstyle=\small\upshape\ttfamily,
  language=SQL,
  morekeywords={PROVENANCE,BASERELATION,INFLUENCE,COPY,ON,TRANSPROV,TRANSSQL,TRANSXML,CONTRIBUTION,COMPLETE,TRANSITIVE,NONTRANSITIVE,EXPLAIN,SQLTEXT,GRAPH,IS,ANNOT,THIS,XSLT,MAPPROV,cxpath,OF,TRANSACTION,SERIALIZABLE,COMMITTED,INSERT,INTO,WITH,SCN,PROV,IMPORT,FOR,JSON,JSON_TABLE,XMLTABLE},
  extendedchars=false,
  keywordstyle=\color{blue},
  mathescape=true,
  escapechar=@,
  sensitive=true,
  stringstyle=\color{Strings},%
  string=[b]'
}

\lstdefinestyle{rsl}
{
tabsize=3,
basicstyle=\small\upshape\ttfamily,
language=C,
morekeywords={RULE,LET,CONDITION,RETURN,AND,FOR,INTO,REWRITE,MATCH,WHERE},
extendedchars=false,
keywordstyle=\color{blue},
mathescape=true,
escapechar=@,
sensitive=true
}

\lstdefinestyle{pseudocode}
{
  tabsize=3,
  basicstyle=\small,
  language=c,
  morekeywords={if,else,foreach,case,return,in,or},
  extendedchars=true,
  mathescape=true,
  literate={:=}{{$\gets$}}1 {<=}{{$\leq$}}1 {!=}{{$\neq$}}1 {append}{{$\listconcat$}}1 {calP}{{$\cal P$}}{2},
  keywordstyle=\color{blue},
  escapechar=&,
  numbers=left,
  numberstyle={\color{green}\small\bf}, 
  stepnumber=1, 
  numbersep=5pt,
}

\lstdefinestyle{xmlstyle}
{
  tabsize=3,
  basicstyle=\small,
  language=xml,
  extendedchars=true,
  mathescape=true,
  escapechar=£,
  tagstyle={\color{blue}},
  usekeywordsintag=true,
  morekeywords={alias,name,id},
  keywordstyle={\color{red}}
}

\lstdefinelanguage{json}{
    basicstyle=\footnotesize\ttfamily,
    numbers=left,
    numberstyle=\scriptsize,
    stepnumber=1,
    numbersep=8pt,
    stringstyle=\color{Strings},%
    showstringspaces=false,
    breaklines=true,
    frame=lines,
    string=[b]"
}


\lstset{style=psql}


\title{Debugging Transactions and Tracking their Provenance with Reenactment}




%
%
%
%

\numberofauthors{3} 

\author{
  \begin{minipage}{1.0\linewidth}
   \centering
    \begin{minipage}{0.2\linewidth}
      \centering
Xing Niu${}^{1}$\\
    \end{minipage}
    \begin{minipage}[t]{0.26\linewidth}
      \centering
Bahareh Sadat Arab${}^{1}$\\
    \end{minipage}
    \begin{minipage}[t]{0.15\linewidth}
      \centering
Seokki Lee${}^{1}$\\
    \end{minipage}
    \begin{minipage}[t]{0.12\linewidth}
      \centering
Su Feng${}^{1}$\\
    \end{minipage}
    \begin{minipage}[t]{0.12\linewidth}
      \centering
Xun Zou${}^{1}$\\
    \end{minipage}\\[2mm]
    \begin{minipage}[t]{0.2\linewidth}
      \centering
Dieter Gawlick${}^{2}$\\
    \end{minipage}
    \begin{minipage}[t]{0.28\linewidth}
      \centering
Vasudha Krishnaswamy${}^{2}$\\
    \end{minipage}
    \begin{minipage}[t]{0.2\linewidth}
      \centering
Zhen Hua Liu${}^{2}$\\
    \end{minipage}    
    \begin{minipage}[t]{0.18\linewidth}
      \centering
Boris Glavic${}^{1}$\\
    \end{minipage}
  \end{minipage}\\
\resizebox{1\linewidth}{!}{
  \begin{minipage}{1.05\linewidth}
    \begin{minipage}[t]{0.5\linewidth}
      \vspace{3mm}
      \centering {\large\bf Illinois Institute of Technology${}^{1}$ }
        \email{\normalsize \{xniu7,barab,slee195,sfeng14,xzou3\}@hawk.iit.edu,bglavic@iit.edu}
    \end{minipage}
    \begin{minipage}[t]{0.51\linewidth}
      \vspace{3mm}
      \centering 
      { \large\bf Oracle Corporation${}^{2}$}
      \email{\normalsize \{dieter.gawlick,vasudha.krishnaswamy,zhen.liu\}@oracle.com}
    \end{minipage}
  \end{minipage}
}
}

\maketitle

\begin{abstract}
Debugging transactions and understanding their execution are of immense importance for developing OLAP applications, to trace causes of errors in production systems, and to audit the operations of a database. However, debugging transactions is hard for several reasons: 1) after the execution of a transaction, its input is no longer available for debugging, 2) internal states of a transaction are typically not accessible, and 3) the execution of a transaction may be affected by concurrently running transactions. We present a debugger for transactions that enables non-invasive, post-mortem debugging of transactions with provenance tracking and supports what-if scenarios (changes to transaction code or data). Using \emph{reenactment}, a declarative replay technique we have developed, a transaction is replayed over the state of the DB seen by its original execution including all its interactions with concurrently executed transactions from the history. Importantly, our approach uses the temporal database and audit logging capabilities available in many DBMS and does not require any modifications to the underlying database system nor transactional workload.
\end{abstract}

\section{Introduction}\label{sec:intro}
 The powerful abstraction of transactions provides a clean and precise semantics for concurrent execution of updates and enables related updates to be grouped together such that their execution as a transaction is atomic. However, developing transactions is hard, because databases lack tools for debugging the concurrent execution of transactions. There are several debuggers for procedural extensions of SQL (e.g., PL/SQL or T-SQL\footnote{
  Microsoft T-SQL Debugger: \url{http://msdn.microsoft.com/en-us/library/cc645997.aspx}\\ RapidSQL: \url{https://www.idera.com/rapid-sql-ide}})
 that provide features typically supported by debuggers for imperative languages, e.g.,  step-wise execution, observing variable values, and manipulating the content of variables. 
However, these debuggers treat SQL statements as black boxes, i.e., they do not expose the dataflow within an SQL statement (e.g., a query or update). 
Furthermore, they do not support debugging a past execution of a transaction within its original environment which is necessary for detecting bugs that are based on concurrency anomalies caused by lower isolation levels and bugs that only materialize for certain states of a database.  
Provenance and declarative debugging techniques~\cite{GR13} can unearth intermediate states of tables and expose data dependencies in declarative languages. However, except for our own work~\cite{AG17,AG14,XN17} 
there are no approaches that support transactions. 
In general, debugging transactions and tracking their provenance are challenging for the following reasons:

\mypar{C1. Debugging may alter the database}
When executing a transaction to debug it, the DML 
statements executed by the transaction will alter the database which is unacceptable for production environments. One way to overcome this problem is to maintain a separate development database for debugging. However, it may be hard to reproduce bugs that are encountered in the production environment unless the development database is kept in sync with the production database. Furthermore, additional work is required to restore the development database to its original state after each debugging session to make debugging repeatable.
 

\mypar{C2. Past database states are transient}
The database state seen by a past transaction is typically not available for post-mortem debugging. That is, many buggy transaction executions will not be detected since the information required to detect them is not available after the fact. 
Furthermore, once a buggy execution is detected, it is not possible to replicate the conditions that lead to the bug. \textit{Time travel} functionality, i.e., providing query access to past versions of a table, is supported by many database systems (e.g., Oracle, MSSQL, and DB2). However, the snapshots returned by time travel only contain committed changes of transactions - intermediate versions of tuples that only existed during the execution of a transaction are not available.


\mypar{C3. Dataflow within SQL statements is opaque}
As mentioned above, current debuggers do not allow the dataflow within an SQL statement to be inspected. Database provenance  provides such dataflow information by recording which input tuples were used to compute an output tuple of an operation and how they have been combined. 
Provenance can also help us to focus the debugging process on data affected by an operation. For example, consider a transaction that withdraws money from a customer's account. Rather than showing the full account table to the user, it would be better to use provenance  to only show rows that were actually affected by the transaction.


\mypar{C4. Non-serializable isolation levels} 
Most databases support several isolation levels (e.g., \lstinline!READ COMMITTED!) in addition to serializable execution to enable users to trade consistency for performance, i.e., 
less strict isolation levels permit certain concurrency anomalies such as write-skews~\cite{BB95} and non-repeatable reads to occur. Furthermore, some da\-ta\-bases (e.g., Oracle and older versions of Postgres) do only support snapshot isolation (SI) which does not guarantee serializability.
The use of lower isolation levels is common in real applications and compensated for by carefully designing applications 
and their transactions to avoid anomalies. 
While this approach can significantly increase performance of transaction processing, it places a high burden on the developer. 
When a transaction produces an unexpected effect, this may be due to a logical bug in its implementation or because the programmer failed to predict that an anomaly could occur. 
Anomalies are hard to debug since they cannot be reproduced by rerunning the transaction in an isolated testing environment unless the transactions that were involved in the anomaly are repeated using precisely the same interleaving of operations as in the original execution.


\begin{figure}[t]
 \centering
\begin{minipage}{1\linewidth}
\begin{minipage}{1.0\linewidth}
\centering
\textbf{Bob's Withdrawal Transaction}\\[-2mm]
\begin{lstlisting}
UPDATE account SET bal = bal - :amount 
WHERE cust = :name AND typ = :type;
\end{lstlisting}$ $\\[-4mm]
\begin{lstlisting}
INSERT INTO overdraft (
 SELECT cust, a1.bal + a2.bal 
 FROM account a1, account a2 
 WHERE a1.cust = :name AND a1.cust = a2.cust 
  AND a1.typ != a2.typ AND a1.bal + a2.bal < 0);
\end{lstlisting}
\vspace*{-1mm}
\textbf{Execution Order of Transactions $T_1$ and $T_2$}\\
\vspace{-3mm}
\begin{center}
  \resizebox{0.8\linewidth}{!}{
    \begin{tikzpicture}[
      scale=7,
      closedint/.style={|-|,thick},
      lopenint/.style={(-],thick},
      ropenint/.style={[-),thick},
      openint/.style={(-),thick}
]
      \draw[|->, thick] (-0.1,0) -- node[below]{\textbf{Time}} (1.1,0); 
      \draw[closedint] (0,0.07) -- node[below]{$\bf T_1$} (0.5,0.07);
      \draw (0,0.05) -- node[above]{update} (0,0.09);
      \draw (0.2,0.05) -- node[above]{insert} (0.2,0.09);
      \draw (0.5,0.05) -- node[above]{commit} (0.5,0.09);
      \draw[closedint] (0.4,0.13) -- node[below]{$\bf T_2$} (1,0.13);
      \draw (0.4,0.11) -- node[above]{update} (0.4,0.15);
      \draw (0.8,0.11) -- node[above]{insert} (0.8,0.15);
      \draw (1,0.11) -- node[above]{commit} (1,0.15);
    \end{tikzpicture}
  }
\end{center}
\vspace*{-3mm}
\textbf{Bind Parameters for Transactions $T_1$ and $T_2$}\\[1mm]
{\scriptsize
\begin{tabular}{|c|c|c|c|}
\thead{Transaction} & \thead{\texttt{:name}}& \thead{\texttt{:amount}} & \thead{\texttt{:type}} \\ 
 $T_1$ & \texttt{Alice} & 70 & Checking \\ 
 $T_2$ & \texttt{Alice} & 40 & Savings \\ \hline
\end{tabular}
}
\end{minipage}


 \end{minipage}\\[1mm]
\caption{Running example transactions}
\label{fig:Transactions Example}
\end{figure}
\begin{figure}[t]
  \centering
  \begin{minipage}{0.28\linewidth}
  {\small  \textbf{(a)} Database before execution of $T_1$ and $T_2$}
  \end{minipage}
  \begin{minipage}{0.46\columnwidth}
    \scriptsize
    \centering { \bf account}\\[1mm]
    \begin{tabular}{|c|c|c|}
      \thead{cust} & \thead{typ} & \thead{bal}   \\ \hline
           Alice & Checking & 50 \\
          Alice & Savings & 30 \\ \hline
    \end{tabular}\\[2mm] 
  \end{minipage}
  \begin{minipage}{0.2\columnwidth}
    \scriptsize
    \centering { \bf overdraft}\\[1mm]
    \begin{tabular}{|c|c|}
      \thead{cust} & \thead{bal}   \\ \hline
    \end{tabular}\\[2mm] 
  \end{minipage}\\[1mm]
  \hrule$ $\\[1mm]
  \begin{minipage}{0.29\linewidth}
  {\small  \textbf{(b)} Database after execution of $T_1$}
  \end{minipage}
  \begin{minipage}{0.46\columnwidth}
    \scriptsize
    \centering { \bf account}\\[1mm]
    \begin{tabular}{|c|c|c|}
      \thead{cust} & \thead{typ} & \thead{bal}   \\ \hline
           Alice & Checking & -20 \\
          Alice & Savings & 30 \\ \hline
    \end{tabular}\\[2mm]
  \end{minipage}
  \begin{minipage}{0.2\columnwidth}
    \scriptsize
    \centering { \bf overdraft}\\[1mm]
    \begin{tabular}{|c|c|}
      \thead{cust} & \thead{bal}   \\ \hline
    \end{tabular}\\[2mm] 
  \end{minipage}\\[1mm]
\hrule$ $\\[1mm]
  \begin{minipage}{0.29\linewidth}
    {\small \textbf{(c)} Database after execution of $T_2$}
  \end{minipage}
  \begin{minipage}{0.46\columnwidth}
    \scriptsize
    \centering { \bf account}\\[1mm]
    \begin{tabular}{|c|c|c|}
      \thead{cust} & \thead{typ} & \thead{bal}   \\ \hline
           Alice & Checking & -20 \\
          Alice & Savings & -10 \\ \hline
    \end{tabular}\\[2mm] 
  \end{minipage}
  \begin{minipage}{0.2\columnwidth}
    \scriptsize
    \centering { \bf overdraft}\\[1mm]
    \begin{tabular}{|c|c|}
      \thead{cust} & \thead{bal}   \\ \hline
    \end{tabular}\\[2mm] 
  \end{minipage}
$ $\\[1mm]
  \caption{Running example database states}
  \label{fig:running-example-instance}
\end{figure}
\begin{exam}\label{ex:query-example}
Bob is a developer at a bank that runs a database using the \emph{snapshot isolation (SI)} concurrency control protocol~\cite{BB95} 
(e.g., Oracle). 
He is tasked with writing a transaction for withdrawing money from a customer's checking or savings account (a table \texttt{account(cust,typ,bal)}). If after the withdrawal the total balance of the checking and savings account for the customer are below 0, then an overdraft record should be inserted into a table \texttt{overdraft(cust,bal)}. Bob implements the transaction shown in Fig.~\ref{fig:Transactions Example} that runs an update followed by an insert using a query that detects overdrafts. After some tests that are uneventful, Bob's solution is deployed. However, it turns out that Bob's transaction does not always report overdrafts correctly. Assume that transactions $T_1$ and $T_2$ as shown in 
Fig.~\ref{fig:Transactions Example} have been executed concurrently with $T_2$ committing last. Fig.~\ref{fig:running-example-instance} shows the database state before and after execution of $T_1$ and $T_2$.
As shown in Fig.~\ref{fig:running-example-instance} (c), these transactions cause an overdraft for Alice that is evident in the database state after $T_2$'s commit (since $-20 + (-10) < 0$).  
However, neither $T_1$ nor $T_2$ have reported this overdraft.
The cause of this problem is that SI does not guarantee serializability. In fact, it can lead to a concurrency anomaly called \textit{write-skew}~\cite{BB95} as exemplified in this example. Under SI, a transaction $T$ runs over a private snapshot of the database that contains changes made by transactions that committed before $T$ started. Thus, $T_1$ and $T_2$ do not see each others changes. Both transactions compute the total balance using an outdated balance for the other account. For instance, $T_2$ sees the previous balance of \$50 for Alice's checking account and the condition of the overdraft check evaluates to $50 + (-10) = 40 \not< 0$.  If a debugger would be available that enables Bob to inspect the versions of these tables seen by the execution of Transaction $T_2$, 
 then he would be able to determine that the problem was caused by reading an outdated balance. 
Afterwards, Bob can fix the problem, e.g., using \emph{promotion} as we will explain further  in Section~\ref{sec:debug-gui}. 
 If Bob has to manually debug the transaction in a development environment, then the error would not materialize unless he interleaves the execution of two transactions for the same customer but different account types. However, this requires that Bob understands that this particular interleaving is likely causing the error.
\end{exam}

Debugging of transactions would be greatly simplified if a debugger would show the intermediate states produced by the past execution of a transaction. It should be possible to trace the provenance of individual tuple versions (which operations of the transaction affected them and which earlier tuple versions were they derived from) to better understand an execution. Furthermore, the user should be able to explore the effect of hypothetical changes to data or transaction statements (\textit{what-if}).
In this paper, we present a novel debugger for transactions that uses reenactment~\cite{AG17,AG14,XN17}, a declarative replay techniques we have developed, to recreate the state of the database observed by the original execution of a transaction including all its interactions with other transactions from the history that executed concurrently. Using reenactment, we can overcome the challenges discussed above. Reenactment uses the time travel and audit logging features available in many DBMS to be able to reconstruct any past database state (C2). Since reenactment works by running queries, debugging does not alter the database (C1). Reenactment supports retroactive provenance tracking (C3) and we have demonstrated that it is possible to reenact transactions executed under non-serializable isolation levels~\cite{AG17} (C4).
%
This makes it possible to, e.g., debug concurrency anomalies.
In the following, we introduce our debugger (Section~\ref{sec:debug-gui}), give a brief overview of reenactment and its implementation in our GProM system~\cite{AG17,AG14} (Section~\ref{sec:trans-reen} and~\ref{sec:gprom-approach}), and give an outline of the demonstration (Section~\ref{sec:demo}). Our debugger does currently not expose procedural extensions of SQL. However, it can still be used to debug the SQL operations of transactions that were executed as stored procedures. Integrating our system with debuggers for procedural languages (e.g., the debuggers mentioned in Section~\ref{sec:intro}) is an interesting avenue for future work.

\section{The Debugger}
\label{sec:debug-gui}
\begin{figure}[t]
 \centering$\,$\\[8.5mm]
\begin{minipage}{1\linewidth}
\includegraphics[width=1\columnwidth,trim=0 112pt 0 180pt]{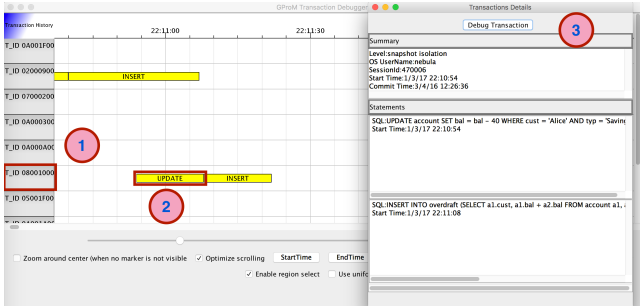}
 \end{minipage}
\caption{Main panel showing a transaction history and a panel with  transaction details (\fignumref{3}).} 
\label{fig:demo-main-gui}
\end{figure}

The main panel of the debugger's GUI (Fig.~\ref{fig:demo-main-gui}) shows a horizontal time line of transactions executed in the past. This panel used to identify suspicious or interesting transaction executions to debug. 
The timeline view is instantiated based on the transactional history of a database by querying the audit log of this database (see Section~\ref{sec:trans-reen} for details).
The user can zoom in and out, restrict the view to a certain time interval, and scroll along the time axis. While currently not supported, it would be straightforward to implement more powerful search functionality, e.g., full text search over SQL commands of transactions. Since finding an execution of interest is orthogonal to debugging it (our main focus) we leave such extensions to future work. 
Each row in the time chart corresponds to a transaction with its identifier shown in the bar on the left (e.g., \fignumref{1}).  
Statements of transactions are shown as intervals (\fignumref{2}). The starting point of such an interval is the time when the statement was executed while the end time is the start time of the next statement or the transaction's
commit time (for the last statement). 
For example, 
\fignumref{1} shows Transaction $T_2$ from Ex.~\ref{ex:query-example}. 

\begin{figure}[t]
 \centering
\begin{minipage}{1\linewidth}
\includegraphics[width=1\columnwidth,trim=0 135pt 0 170pt]{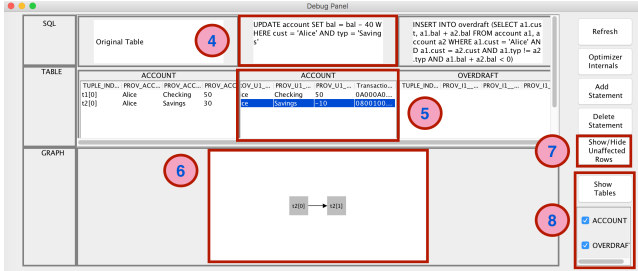}
 \end{minipage}
\caption{Debug panel showing all intermediate states of relations. Clicking on a tuple $t$ shows its provenance graph (\fignumref{6}). Modifying the SQL code of a statement or the content of a table instructs the system to create a what-if scenario.}
\label{fig:demo-debug-screen}
\end{figure}

Selecting a transaction opens up a panel (\fignumref{3}) which provides additional information about this transaction such as isolation level, commit time, name of the user executing the transaction, session ID, and the SQL code and start time for each statement of the transaction. 
Once a user has identified a set of interesting transaction executions, she can inspect their details  (\fignumref{3}) and 
select them for in depth debugging.


\mypar{Debugging}
The user can click the ``Debug Transaction'' button in the detail panel (\fignumref{3}) to debug a selected transaction. The debug panel (Fig.~\ref{fig:demo-debug-screen}) shows one column for each operation of the transaction plus a column for the initial states of the relations accessed by the transaction. Each such column shows the SQL code of the statement (\fignumref{4}) and 
the table (\fignumref{5}) modified by the statement (the version created by the statement). For each tuple version, we show which transaction created that version.
When a user clicks on a tuple version 
we generate a provenance graph (\fignumref{6}) which shows all past tuple versions involved in the creation of this tuple (e.g., the previous versions of a tuple modified by an update). Each node in such a graph represents a tuple version and edges denote derivation. 
The default for this panel is to show only rows affected by at least one statement of the transaction (plus rows in the provenance of such rows) to limit the presentation to what is relevant for debugging. However, if the user is interested in seeing all rows of these tables, he can click on the ``Show/Hide Unaffected Rows'' button (\fignumref{7}). Furthermore, he can select which tables should be shown in the interface (\fignumref{8}).\\[-2mm]

\begin{exam}\label{ex:bob-debugs}
Continuing with Ex.~\ref{ex:query-example}, Bob investigates the missing overdraft record and uses the debugger to inspect transactions $T_1$ and $T_2$. Fig.~\ref{fig:demo-debug-screen} shows a screenshot of the debug panel for  $T_2$. 
He observes that both transactions did not insert any tuples into the overdraft table and the insert statement of $T_2$ sees an outdated balance (50 instead of -20) for the checking account which is why the transaction did not create an overdraft record. Thus, he has identified the write-skew anomaly that caused the missing overdraft.   
\end{exam}



\myparr{What-if scenarios}
Our system supports two types of what-if scenarios: 1) the user can edit the data in a table and 2) the user can modify, delete, or add an update statement. If the user edits a table $R$, we create a temporary table storing the updated version of table $R$ (say $R'$). 
We, then, replace all accesses to $R$ with $R'$ in the reenactment query and reevaluate it. 
To support the second type of functionality, we reconstruct the reenactment query using the modified statements instead of the original statements and reevaluate this query to refresh the display. 
For example, 
Bob could use this functionality to add a redundant update \lstinline!UPDATE account SET bal = bal WHERE cust = :name! to his tra\-ns\-ac\-tion that updates both the savings and checking account of a customer. This trick, often called \textit{promotion}, guarantees that no two concurrent executions of Bob's transaction can update accounts of the same customer. In the example, this would force $T_2$ to abort. 




\section{Transaction Reenactment}
\label{sec:trans-reen}

In~\cite{AG17,AG14}, we have introduced reenactment, a declarative technique
for replaying parts of a \textit{SI} transactional history\footnote{We support isolation levels \lstinline!SERIALIZABLE! and \lstinline!READ  COMMITTED! for systems that run SI.} using queries.
SI is a widely applied multi-versioning concurrency control protocol that is supported by,
e.g., Oracle, PostgreSQL, and MSSQL.
  We have proven~\cite{AG17} 
that a reenactment query for a transactional history (or parts thereof)
 produces the same result (updated tables) and has the same provenance (data
dependencies) as the original history. 
A reenacted transaction precisely simulates the original execution of the transaction including all its interactions with concurrently running transactions. 
Thus, reenactment can be used to retroactively capture the
provenance of a past transaction by constructing its reenactment query,
instrumenting this query for provenance tracking, and evaluating it over
the database state seen by the transaction using time travel. For instance, this enables debugging of concurrency anomalies as outlined above.
To construct
a reenactment query for a transaction, we need to know the SQL code of updates
run by the transaction and, for each update, when it was executed.
A query-able audit log of executed SQL statements supported by
many databases (e.g., Oracle and DB2)
provides
sufficient information to enable reenactment. Importantly, reenactment queries instrumented for provenance tracking can be expressed in SQL. Thus, they can be executed using any database that supports audit logging and time travel.\footnote{For systems that do not support these features, it is possible to use 
  triggers to implement them.} 
Based on our experience~\cite{AG17} 
with commercial DBMS X\footnote{Name omitted due to licensing restrictions.}, activating these features results in moderate overhead (20\% for write-only workloads and about 5\% for mixed workloads). 
However, since many users already use these features for other purposes (e.g., auditing), our approach does not result in any  additional overhead for these users.
We also support  reenacting a prefix of a transaction which is useful for restoring the state of a table seen by a particular statement within a transaction. Furthermore, we can use reenactment to support what-if scenarios by reenacting a modified version of a past transaction. While the details of reenactment are beyond the scope of this paper (see~\cite{AG17,AG14}), we further illustrate the idea by example.

\begin{exam}\label{ex:reenact-example}
Consider the update of 
$T_1$. To reproduce the updated version of relation \texttt{account} produced by $T_1$, we construct a reenactment query that simulates the update. An SQL update returns the updated versions of tuples that fulfill the update's condition and the original version of all tuples that do not fulfill the condition. We compute this relation
using  \lstinline!CASE! to decide whether to update a tuple or not.
Since the update is the first operation of $T_1$, this version consists of committed changes by transactions executed before $T_1$ and, thus, can be accessed using time travel. The modifications to attribute values expressed in the \lstinline!SET! clause of the update can be expressed as a projection (\lstinline!SELECT! clause). The update of $T_1$ can be reenacted as shown below. Here, \lstinline!AS OF! denotes using time travel to get back a past version of a table (assuming the update was executed at \textsf{\upshape '2016-03-01'}).
\begin{lstlisting}
SELECT cust, typ, 
   CASE WHEN cust = 'Alice' AND typ = 'Checking' 
        THEN bal - 70 ELSE bal END AS bal 
FROM account AS OF '2016-03-01'
\end{lstlisting}
\end{exam}

\section{The GProM Approach}
\label{sec:gprom-approach}

We have implemented reenactment and on demand provenance tracking techniques in our
GProM system~\cite{AG14}. GProM is a database independent middleware (support for new
backends can be added through plugins) that 
supports an SQL dialect with language constructs for requesting the provenance of
a query or transaction. Provenance requests return standard relations and are
treated as queries, e.g., a provenance request can be part of
a more complex SQL query.

\begin{figure}[t]
 \centering
\begin{minipage}{1\linewidth}
\includegraphics[width=1\columnwidth]{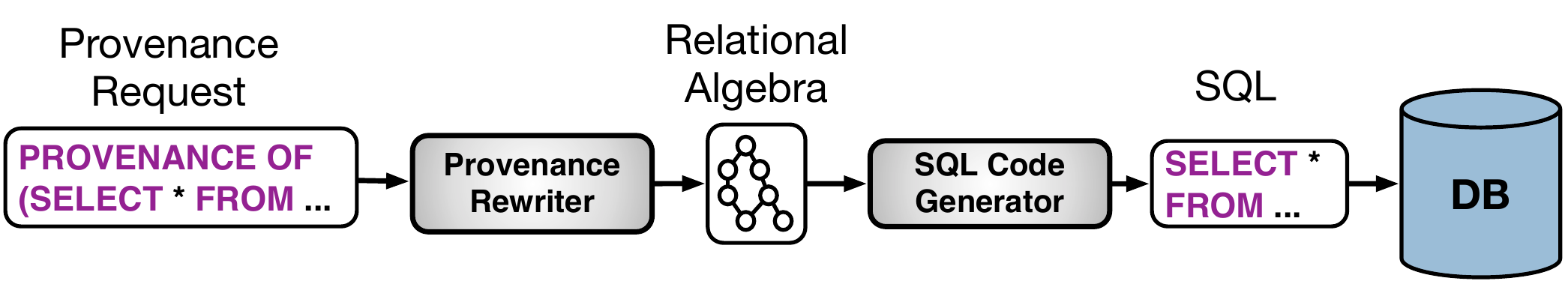}
 \end{minipage}
\caption{Processing provenance queries with GProM.}
\label{fig:general-rewrite-approach}
\end{figure}


\mypar{GProM Query Processing}
 Fig.~\ref{fig:general-rewrite-approach} gives a brief overview of how queries 
 are processed by GProM. The user submits an SQL query, potentially including one or more provenance requests. 
 The \textit{parser} and \textit{analyzer} modules translate this extended SQL query into a relational algebra graph (used as an intermediate language by GProM). The \textit{provenance rewriter} then adds any necessary instrumentation for provenance tracking and uses the \textit{reenactor} to construct reenactment queries if necessary. The output of the provenance rewriter is a  relational algebra expression that no longer contains any provenance-specific operators. This expression is then translated into the SQL dialect of the backend database using a database specific \textit{SQL code generator} plugin. The resulting SQL query is 
sent to the backend database and its results are passed on to the user.  
By applying provenance-specific optimizations~\cite{AG17,XN17} we can reenact complex transactions over tables with millions of rows within seconds. 
\section{Demonstration Overview}
\label{sec:demo}

We will bring a laptop running the debugger and a VM with the backend database. We will prepare a transaction history that contains simple examples such as the ones shown in the paper as well as more complex transactions showcasing various anomalies (e.g., \textbf{write-skew} and \textbf{non-re\-peat\-a\-ble reads}). Attendees will also be able to execute new transactions and we will adapt the presentation based on their individual background and interests. 

\section{Conclusions}
\label{sec:clu}

We introduce 
a debugger for transactions that allows  users to inspect intermediate states of relations produced by a past execution of a transaction, to trace dataflow among tuples (i.e., a tuple's provenance), and to explore the effect of hypothetical changes to the data or SQL statements executed by a transaction. 
This debugger uses the reenactment capabilities of our GProM system to replay transactional workloads. GProM implements  
reenactment as SQL queries using the temporal and auditing features available in many databases. 




{\small
\bibliographystyle{abbrv}
\bibliography{gprom_demo}
}

\end{document}